\documentstyle[12pt,epsfig]{article}

\newlength{\dinwidth}                                                      
\newlength{\dinmargin}                                                      
\def\lapproxeq{\lower .7ex\hbox{$\;\stackrel{\textstyle                                                      
<}{\sim}\;$}}                                                      
\def\gapproxeq{\lower .7ex\hbox{$\;\stackrel{\textstyle                                                      
>}{\sim}\;$}}

\def\bea{\begin{eqnarray}}                                                      
\def\eea{\end{eqnarray}}                                                      
\def\Qtbold{\mbox{\boldmath${Q}$}_t}
\def\qbold{\mbox{\boldmath${q}$}}

\begin{document}                                                      
\titlepage                                                      
\begin{center}                                                      
{\LARGE \bf  
Unintegrated gluon distributions from the transverse coordinate representation 
of the  CCFM equation in the single loop approximation}\\ 
\vspace*{0.4in}   
J. \ Kwieci\'nski\\   

\vspace*{0.5cm}   
 
{\it H. Niewodnicza\'nski Institute of Nuclear Physics,   
 Krak\'ow, Poland} \\   
\end{center}
\begin{abstract}
We utilise the fact  that the CCFM equation in the single loop approximation can be diagonalised by 
the Fourier-Bessel transform.  The analytic 
solution of the CCFM equation for the  moments $f_{\omega}(b,Q)$ of the scale dependent 
gluon distribution is 
obtained, where $b$ is the transverse coordinate conjugate to the transverse momentum  
of the gluon.   The unintegrated gluon distributions obtained from this  
solution are analysed.    
It is shown how the  approximate treatment of the exact solution makes it possible to 
express  the unintegrated gluon distributions in terms of  the integrated ones.    
The corresponding approximate expressions for the unintegrated gluon distribution 
are compared with exact 
solution of the CCFM equation in the single loop approximation. 
\end{abstract}

\newpage

\section{Introduction}
The basic, universal quantities of the QCD improved parton model are the scale 
dependent parton distributions, like the gluon distribution  $g(x,Q^2)$, 
where $x$ denotes the momentum fraction.  The (integrated) parton 
distributions can be related to the less inclusive 
distributions $f(x,Q_t,Q)$ unintegrated over transverse momentum $Q_t$ of the parton.  
Those {\it uninegrated} distributions are often needed in  less inclusive measurements 
which are sensitive to the transverse momentum of the parton \cite{DDT} - \cite{KHMR}.  
%It should also be emphasised that 
%at low $x$ it is the unintegrated 
%distributions which are somehow more fundamental than the integrated ones.\\

The unintegrated, scale dependent distributions are described in QCD by the 
Catani-Ciafaloni-Fiorani-Marchesini (CCFM) equation \cite{CCFM} - \cite{JUNGS} 
based upon quantum coherence which implies 
angular ordering \cite{DKTM}.  Its very nice feature is 
 the fact that it embodies both the DGLAP and BFKL evolutions at low $x$.  
In the region of large and moderately small values of $x$, where the small $x$ 
effects  can be neglected the CCFM equation becomes equivalent to the 
(LO) DGLAP evolution.  
This approximation corresponds to the so called 'single loop' approximation \cite{BRW,GMBRW}. \\

The CCFM equation interlocks in a rather complicated way the two 
relevant scales i.e.  the transverse momentum $Q_t$ of the parton 
and the hard scale $Q$.   
 The main purpose of this paper is to explore the fact that in the 'single loop' 
approximation the CCFM equation can be solved exactly in the transverse coordinate 
representation conjugate to the transverse momentum of the parton.  The unintegrated 
distributions can then be obtained from the Fourier-Bessel transform of this solution. 
  Although the 'single loop' approximation neglects  small $x$ 
effects and so it is not valid at very small $x$ it can be a reasonable approximation 
 at large and moderately small values of $x$ ( $x\ge 0.01$ or so), which is certainly the region 
of phenomenological interest \cite{KMR1,KHMR}. 
Analytic insight into the 
{\it exact} solution of the CCFM equation in the single loop approximation will also  make it possible to critically  
examine and justify approximate formulas linking the unintegrated distributions 
to the integrated ones in the region where the LO DGLAP dynamics should be adequate 
\cite{DDT,KMR1,KHMR}.\\        

 The content of our paper is as follows:  In the next Section we recall the 
CCFM equation for the unintegrated gluon distribution.   In Section 3 we discuss  the 'single loop' approximation of 
this equation in the transverse coordinate 
representation.  We show that it can be solved exactly for the moment function 
$f_{\omega}(b,Q)$, where $b$ denotes the transverse coordinate conjugate to the 
transverse momentu $Q_t$ of the gluon. 
We do also show how the approximate forms of this solution expressing the unintegrated distributions 
in terms of the integrated ones \cite{DDT,KMR1}   originate from the exact solution. 
 In Section 4 we present numerical results 
for the unintegrated gluon distributions based on the solution 
of the CCFM equation in the transverse coordinate representation.  We also confront 
exact solution with its approximate forms.  
Finally in Section 5 we summarise  our main results 
and give our conclusions.       
\section{The CCFM equation}
Parton cascade with angular ordering generates  the  Catani, Ciafaloni, Fiorani, Marchesini  
(CCFM) equation \cite{CCFM} for the unintegrated, scale dependent gluon distribution 
$f(x,Q_t,Q)$ 
in the proton, where $x,Q_t$ and $Q$ denote the longitudinal momentum fraction  
carried by the gluon, transverse momentum of the gluon and the 
hard scale respectively.  The latter is specified by the maximal value of the emission 
 angle.  
The CCFM equation has the following form:   

\vspace*{1cm}

$$
f(x,Q_t,Q) =\bar f^0(x,Q_t,Q) + \int{d^2\qbold \over \pi q^2} \int_x^1 {dz\over z}  
\Theta(Q-qz) \Theta(q-q_0)
{ \alpha_s\over 2\pi } \Delta_s (Q,q,z)
$$

\begin{equation}
 \left[2N_c\Delta_{NS}(Q_t,q,z) + 
{2N_cz\over (1-z)} + 
z\bar P_{gg}(z)\right]
 f\left({x\over z},|\Qtbold+(1-z)\qbold|,q)\right)
\label{ccfm1}
\end{equation}
where $\Delta _S(Q,q,z)$ and $\Delta_{NS}(Q_t,q,z)$ are the Sudakov and non-Sudakov 
form factors.  They are given by the following expressions:  

\begin{equation}
\Delta _S(Q,q,z)=exp\left[-\int_{(qz)^2}^{Q^2}{dp^2\over p^2}{\alpha_s\over 2 \pi}
\int _0^{1-q_0/p}dzzP_{gg}(z)\right]
\label{ds}
\end{equation}

\begin{equation}
\Delta_{NS}(Q_t,q,z)= exp\left[-\int_z^1{dz'\over z'} 
\int_{(qz')^2}^{Q_t^2}
{dp^2\over p^2}{2N_c\alpha_s\over 2 \pi}\right]
\label{dns}
\end{equation}

For simplicity we neglect  possible  quark contributions.  The function $\bar P_{gg}(z)$ 
is:
\begin{equation}
\bar P_{gg}(z)=2N_c[-2+z(1-z)]
\label{barp}
\end{equation}
and corresponds to the non-singular part of the $g\rightarrow gg$ splitting function 
$P_{gg}(z)$
\begin{equation}
P_{gg}(z)=2N_c\left[{1\over z}+{1\over 1-z}\right]+\bar P_{gg}(z)
\label{pz}
\end{equation} 
The argument of $\alpha_s$ will be specified later.\\

In principle the CCFM equation has been obtained using only the singular parts of the 
splitting function proportional to $1/z$ and $1/(1-z)$.  We add the non singular part
$\bar P_{gg}(z)$ to the kernel of this equation in order to obtain the complete DGLAP evolution 
in  the 'single loop' approximation.          
The two-dimensional vector $\qbold$ in equation (\ref{ccfm1}) is related to the transverse momentum 
$\qbold_t$ of the emitted gluon

\begin{equation}
\qbold_t=(1-z)\qbold
\label{q}
\end{equation}
The constraint $Q>qz$ reflects the angular ordering and the inhomogeneous term 
$\bar f^0(x,Q_t,Q)$ is related to the input non-perturbative gluon distribution. 
It also contains effects of both the Sudakov and non-Sudakov form-factors.

It should be observed that if the cut-off $qz'$ in the definition of the 
non-Sudakov form-factor is replaced by the fixed cut-off $q_0$ then the non-Sudakov 
form-factor reduces to the form-factor reflecting the reggeisation of the gluon, i.e.:

$$
qz' \rightarrow q_0 \rightarrow 
\Delta_{NS} = exp [(2\alpha_G(Q_t^2)-2)ln(1/z)]
$$ 

$$\alpha_G(Q_t^2)=1-\int_{q_0^2}^{Q_t^2}
{dp^2\over p^2}{N_c\alpha_s\over 2 \pi}
$$

The unintegrated scale dependent gluon distribution $f(x,Q_t,Q)$ 
is related in the following 'standard' way to the 
conventional (scale dependent) integrated gluon distribution $xg(x,Q^2)$:  

\begin{equation}
xg(x,Q^2) = \int^{Q^2} dQ^2_t f(x,Q_t,Q)
\label{igluon}
\end{equation}

In the  'single loop' approximation of the CCFM equation (\ref{ccfm1}) the angular ordering 
constraint  
$\Theta(Q-qz)$ is replaced by $\Theta(Q-q)$ and the non-Sudakov form-factor 
$\Delta_{NS}$ is set equal to  unity \cite{BRW,GMBRW}.    
Equation (\ref{ccfm1}) then reads: 

$$
f(x,Q_t,Q) =\bar f^0(x,Q_t,Q) + \int{d^2\qbold \over \pi q^2} \int_x^1 {dz\over z}  
\Theta(Q-q) \Theta(q-q_0) 
{ \alpha_s\over 2\pi } \Delta_s (Q,q,z=1)*
$$

\begin{equation}
 \left[2N_c + 
{2N_cz\over (1-z)} + 
z\bar P_{gg}(z)\right]
 f\left({x\over z},|\Qtbold+(1-z)\qbold|,q)\right)
\label{ccfm2}
\end{equation}
It is useful to 'unfold' the Sudakov form factor in equation (\ref{ccfm2}) in order to 
treat  the real emission  and virtual  corrections terms  on equal footing. 
 Unfolded CCFM equation in the single loop approximation takes the following form: 

$$
f(x,Q_t,Q) = f^0(x,Q_t) + \int{d^2\qbold \over \pi q^2}\Theta(q-q_0)
{ \alpha_s(q^2)\over 2\pi}
\int_0^1 {dz\over z}zP_{gg}(z)*
$$

\begin{equation}
\left[
 \Theta(Q-q)\Theta(z-x)
  f\left({x\over z},|\Qtbold+(1-z)\qbold|,q)\right)-z\Theta(Q-q)
f(x,Q_t,q)\right]
\label{ccfmsl}
\end{equation}
 The inhomogeneous term $f^0(x,Q_t)$ is  equal to the input non-perturbative 
gluon distribution in $x$ and $Q_t$.\\

\section{Transverse coordinate representation of the CCFM equation in the single loop 
approximation}

It can easily be  observed that the CCFM equation in the single loop approximation 
(\ref{ccfmsl}) can be diagonalised by the Fourier-Bessel transform: 
\begin{equation}
f(x,Q_t,Q)=\int_0^{\infty} db b J_0(Q_tb) \bar f(x,b,Q)
\label{fb1}
\end{equation}
with the function $\bar f(x,b,Q)$ given by: 
\begin{equation}
\bar f(x,b,Q)=\int_0^{\infty} dQ_t Q_tJ_0(Q_tb)  f(x,Q_t,Q)
\label{fb2}
\end{equation}
where $J_0(u)$ is the Bessel function.  
From equations (\ref{igluon}) and (\ref{fb2}) we get: 
\begin{equation}
\bar f(x,b=0,Q)={1\over 2} xg(x,Q^2)
\label{fgrel}
\end{equation}
The corresponding equation for $\bar f(x,b,Q)$, which follows from equation (\ref{ccfmsl}) 
after taking the Fourier-Bessel transform of both sides of this equation 
reads:
$$
\bar f(x,b,Q) = \bar f^0(x,b) + 
\int_{q_0^2}^{Q^2}{dq^2 \over q^2}{ \alpha_s(q^2)\over 2\pi}\int_0^1 {dz\over z}zP_{gg}(z)*
$$

\begin{equation}
\left\{\Theta(z-x)
 J_0[bq(1-z)q]\bar f\left({x\over z},b,q)\right)-z
\bar f(x,b,q)\right\}
\label{ccfmslb}
\end{equation}
where we put $q^2$ as the argument of $\alpha_s$. This choice of scale gives standard 
LO DGLAP equation for the integrated gluon distribution  for $xg(x,Q^2)=2
\bar f(x,b=0,Q)$. \\
   
 In order to solve equation (\ref{ccfmslb}) it is useful to introduce the moment 
function $\bar f_{\omega}(b,Q)$ 
 \begin{equation}
\bar f_{\omega}(b,Q)=\int_0^1 dx x^{\omega-1}\bar f(x,b,Q)
\label{momfb}
\end{equation}
Equation (\ref{ccfmslb}) implies the following equation for the moment function 
$\bar f_{\omega}(b,Q)$: 

$$
\bar f_{\omega}(b,Q) = \bar f^0_{\omega}(b) + 
\int_{q_0^2}^{Q^2}{dq^2 \over q^2}{ \alpha_s(q^2)\over 2\pi}\int_0^1 dz zP_{gg}(z)*
$$

\begin{equation}
\left\{
 z^{\omega-1}
 J_0[bq(1-z)q]\bar f_{\omega}(b,q)-
\bar f_{\omega}(b,q)\right\}
\label{ccfmmslb}
\end{equation}
The solution of this equation reads:
\begin{equation}
\bar f_{\omega}(b,Q)=f_{\omega}^0(b)exp[S_{\omega}(b,Q)]
\label{sol1}
\end{equation}

where

\begin{equation}
S_{\omega}(b,Q)=\int_{q_0^2}^{Q^2} {dq^2\over q^2}{\alpha_s(q^2)\over 2\pi}\int_0^1 dz 
zP_{gg}(z)\{z^{\omega-1} 
J_0[(1-z)bq]
-1\}
\label{somega}
\end{equation}

At small values of $b$ (i.e. $b<<1/q_0$) we can neglect $b$ dependence 
in $\bar f_{\omega}^0(b)$ and set 
$$\bar f_{\omega}^0(b) \simeq  \bar f_{\omega}^0(b=0)$$ 

We can identify 
$\bar f_{\omega}^0(b=0)$ with the moment of the input (non-perturbative) 
integrated distribution, i.e.
\begin{equation}
\bar f_{\omega}^0(b=0)={1\over 2}g_{\omega}^0
\label{fgb0}
\end{equation}
where 
\begin{equation}
 g_{\omega}^0=\int dQ_t^2 \int_0^1dx x^{\omega-1}f^0(x,Q_t)
\label{gom0}
\end{equation}
We note that at $b=0$ solution ({\ref{sol1}) reduces to the solution of the DGLAP equation 
for the moment function $g_{\omega}(Q^2)$ of the integrated gluon distribution 
$g(x,Q^2)$, i.e.  
\begin{equation}
g_{\omega}(Q^2)= \int_0^1dx x^{\omega}g(x,Q^2)
\label{gmom}
\end{equation} 
To be precise we get 
$$
f(b=0,Q)={1\over 2}g_{\omega}(Q^2)
$$
\begin{equation}
g_{\omega}(Q^2)=g_{\omega}^0exp\left\{\int_{q_0^2}^{Q^2} {dq^2\over q^2}{\alpha_s(q^2)\over 2\pi}\int_0^1 dz 
zP_{gg}(z)[z^{\omega-1} 
-1]\right\}
\label{rggmom}
\end{equation}

It is useful  to rearrange solution (\ref{sol1}) as below: 
\begin{equation}
\bar f_{\omega}(b,Q)=\tilde f_{\omega}(b,Q)T_g(b,Q)
\label{tft}
\end{equation}
where 
\begin{equation}
 \tilde f_{\omega}(b,Q)=\bar f^0_{\omega}(b)
exp\left\{\int^{Q^2} {dq^2\over q^2}{\alpha_s(q^2)\over 2\pi}\int_0^1 dz 
zP_{gg}(z)J_0[(1-z)bq][z^{\omega-1} 
-1]\right\}
\label{tf}
\end{equation}
and the Sudakov-like form-factor $T_g(b,q)$ is given by:
\begin{equation}
T_g(b,Q)=exp\left\{\int_{q_0^2}^{Q^2} {dq^2\over q^2}{\alpha_s(q^2)\over 2\pi}\int_0^1 dz 
zP_{gg}(z)[J_0[(1-z)bq]-1]\right\} 
\label{tg}
\end{equation}

In order to obtain more insight into the structure of the unintegrated 
distribution which follows from the CCFM equation in the single loop approximation 
it is useful to adopt the following approximation of the Bessel function:
\begin{equation}
J_0(u) \simeq \Theta(1-u)
\label{apprj0}
\end{equation}
Using this approximation in equation (\ref{fb2}) we get: 
\begin{equation}
f(x,Q_t,Q)\simeq 2 {d\bar f(x,b=1/Q_t,Q)\over dQ_t^2}
\label{derrel}
\end{equation}

It may be useful to analyse solution (\ref{sol1}) 
using approximation (\ref{apprj0}) which gives 

\begin{equation}
\bar f_{\omega}(b,Q) \simeq {g_{\omega}^0\over 2}T_g(1/b,Q)exp[S^r_{\omega}(b,Q) +\Delta S^r_{\omega}(b,Q)]
\label{barfa}
\end{equation}

where
\begin{equation}
S^{r}_{\omega}(b,Q)\simeq \int_{q_0^2}^{min(1/b^2,Q^2)}
 {dq^2\over q^2}{\alpha_s(q^2)\over 2\pi}\int_0^1 dz zP_{gg}(z) 
[z^{\omega-1} -1]
\label{srdef}
\end{equation}

and 

\begin{equation}
\Delta S^{r}_{\omega}(b,Q)=\int_{1/b^2}^{Q^2} {dq^2\over q^2}{\alpha_s(q^2)\over 2\pi}
\int_{1-1/(bq)}^1 dz zP_{gg}(z) 
[z^{\omega-1}-1]
\label{delsr}
\end{equation}

\begin{equation}
T_g(b,Q) \simeq exp(S^0(b,Q))
\label{tg1}
\end{equation}

where $S^0(b,Q))$ is given by: 
\begin{equation}
S^0(b,Q)=-\int_{1/b^2}^{Q^2} {dq^2\over q^2}{\alpha_s(q^2)\over 2\pi}\int_0^{1-1/(bq)} 
dz zP_{gg}(z) 
\label{s0}
\end{equation}
It may be seen that the form-factor $T_g(1/b,Q)$ given by equations (\ref{tg1},\ref{s0}) has the structure of the 
Sudakov form-factor.  It can also be  seen that the factor
$g_{\omega}^0exp[S^r_{\omega}(b,Q)]$ 
in equation (\ref{barfa}) with $S^r_{\omega}(b,Q)$ defined by equation (\ref{srdef}) 
can be identified with the moment function $g_{\omega}(\mu^2)$ of the integrated gluon distribution at the scale 
$\mu^2=min(1/b^2,Q^2)$.   Neglecting the term  $\Delta S^{r}_{\omega}(b,Q)$ we get:

\begin{equation}
\bar f_{\omega}(b,Q)\simeq T_g(b,Q)g_{\omega}(min(1/b^2,Q^2))
\label{sprll}
\end{equation}
that gives: 

$$
f_{\omega}(Q_t,Q) \simeq 2{\partial \bar f_{\omega}(b=1/Q_t,Q) \over \partial Q_t^2} \simeq 
 $$
\begin{equation}
{\partial [T_g(b=1/Q_t,Q)g_{\omega}(Q_t^2)] \over \partial Q_t^2} 
 \label{ddt}
\end{equation}
for $Q_t<Q$,  and 

$$
f_{\omega}(Q_t,Q)=0
$$
for $Q_t > Q$.  Equation (\ref{ddt}) gives: 
\begin{equation} 
f(x,Q_t,Q) \simeq  {\partial [T_g(b=1/Q_t,Q)xg(x,Q_t^2)] \over \partial Q_t^2} 
\label{ddtx}
\end{equation}
It should be noted that equation (\ref{ddt}) correctly reproduces 
the double logarithmic effects in the region $Q_t<<Q$ \cite{DDT}.  
The formalism presented above is similar to that used for the description of the 
$p_T$ distributions in (for instance) Drell-Yan process (see e.g. 
\cite{PT}).  \\

Taking approximately into account  the remaining contribution in equation (\ref{barfa}) 
gives: 

\begin{equation}
f(x,Q_t,Q) \simeq 
{T_g(b=1/Q_t,Q)\over Q_t^2}\int_0^{1-Q_t/Q}dz  P_{gg}(z)
{\alpha_s\left({Q_t^2\over (1-z)^2}\right)\over 2 \pi}
\Theta(z-x){x\over z}
g\left({x\over z},\left({Q_t\over 1-z}\right)^2\right)
\label{fkmr}
\end{equation}
 Derivation of equation (\ref{fkmr}) is sketched 
in the Appendix.  \\

After replacement $Q_t^2/(1-z)^2 \rightarrow Q_t^2$ in the argument of $\alpha_s$ 
and in the gluon distribution $g(x/z,\mu^2)$ which introduces 
subleading effects, expression (\ref{fkmr}) coincides  
with the representation  used in ref. \cite{KMR1} (modulo subleading terms in the definition 
of the Sudakov form-factor):
\begin{equation}
f(x,Q_t,Q) \simeq 
{T_g(b=1/Q_t,Q)\over Q_t^2}\int_0^{1-Q_t/Q}dz  P_{gg}(z)
{\alpha_s(Q_t^2)\over 2 \pi}
\Theta(z-x){x\over z}
g\left({x\over z},Q_t^2 \right)
\label{fkmr1}
\end{equation}
\section{Numerical results}
In the previous Section we have shown that the CCFM equation in the single loop approximation 
can be solved analytically in the $b$ space, where $b$ is the transverse coordinate 
conjugate to the transverse momentum $Q_t$ of the gluon.  We have also indicated 
approximations which make it possible to relate the unintegrated distributions to the unintegrated 
ones.  In this Section we present results of the numerical analysis of the 
exact solution of the CCFM equation utilising its diagonalisation in the 
transverse coordinate representation.  We shall also confront this exact solution 
with the approximate expressions defined by equations (\ref{ddtx}) and (\ref{fkmr1}).  \\

To this aim we solved equation (\ref{ccfmslb}) for the distribution 
$\bar f(x,b,Q)$ 
and computed the unintegrated distribution $f(x,Q_t,Q)$ from the 
 Fourier-Bessel transform: 
\begin{equation}
f(x,Q_t,Q)=\int_0^{\infty}db b J_0(bQ_t)\bar f(x,b,Q)
\label{fbi2}
\end{equation} 
We started from the input distribution $\bar f^0(x,b)$ 

$$
\bar f^0(x,b)={g_0(x)\over 2}exp(-b^2q_0^2/4)
$$
\begin{equation}
g_0(x)=3(1-x)^5
\label{fxb0}
\end{equation}
where we have set $q_0=1 GeV$.  In Fig. 1 we plot the function $Q_t^2 f(x,Q_t,Q)$ 
as the function of $Q_t$ for $Q^2=100 GeV^2$ and for two valus of $x$, 
$x=0.01$ (upper curve) and $x=0.1$ (lower curve).  
In Figure 2 and 3 we compare those exact solution with approximate expressions 
(\ref{ddtx}) and (\ref{fkmr1}). We find that equation (\ref{fkmr1}) gives somehow 
better approximation of the exact solution except for the 
'end points' $Q_t^2 \sim Q_0^2$ and $Q_t^2 \sim Q^2$.  The simple 
formula (\ref{ddtx}) is a reasonable approximation of the exact solution for small values of $Q_t$. 
It may however give negative contribution at large $x$($x \sim 0.1$) and large $Q_t^2$.  

\section*{Summary and conclusions}
In this paper we have utilised  the transverse  coordinate representation   
of the CCFM equation   
in order to get an analytical insight into its solution.  The transverse coordinate 
 representation 
has been widely used for the discussion of the soft gluon resummation effects in 
the transverse momentum distribution of Drell - Yan pair etc., \cite{PT}.  In our paper we have 
utilised the fact that  this representation diagonalises the CCFM equation in the single 
loop approximation and can be very helpful for obtaining   
 unintegrated parton distribution satisfying the CCFM equation in this approximation. 
We have shown that the CCFM equatiion in the single loop approximation can be solved analytically 
for the moment function $f_{\omega}(b,Q)$, where $b$ is the transverse coordinate 
conjugate to the transverse momentum of the gluon.  We have also confronted the unintegrated 
gluon distributions with approximate expressions which were discussed in the literature.  
The single loop approximation neglects small $x$ effects in the CCFM equation and, in particular, 
it neglects virtual corrections responsible for the non-Sudakov form-factor.   
 This form-factor  
generates contributions which are no longer diagonal in the $b$ space  
and so the merit of using this representation beyond the single-loop approximation 
is less apparent.   However, in the leading $ln(1/x)$ approximation 
at small  $x$ the  CCFM equation  reduces  to  the BFKL equation 
with no scale dependence and 
the kernel of the BFKL 
equation in the $b$ space is the same as the BFKL kernel in the (transverse) 
momentum space.  One can expect that the transverse coordinate representation of the 
CCFM equation may eventually appear
 to be  helpful  beyond the 
single loop and BFKL approximations.  
\section*{Acknowledgments} 
I thank Krzysztof Golec-Biernat for useful discussions. 
This research was partially supported
by the EU Fourth Framework Programme `Training and Mobility of Researchers',
Network `Quantum Chromodynamics and the Deep Structure of Elementary
Particles', contract FMRX--CT98--0194 and  by the Polish 
Committee for Scientific Research (KBN) grants no. 2P03B 05119 and 5P03B 14420.

\section*{Appendix}

%%%%%%%%%%%%%%%%%%%
 In this Appendix we derive equation (\ref{fkmr}).  To this aim we start from the following 
improved approximation of the derivative $\partial \bar f_{\omega}(b=1/Q_t,Q)/\partial Q_t^2$: 
\begin{equation}
2{\partial \bar f_{\omega}(b=1/Q_t,Q)\over \partial Q_t^2} \simeq 
{\partial [T_g(b=1/Q_t,Q)g_{\omega}(Q_t^2)]\over \partial Q_t^2}+
  T_g(b=1/Q_t,Q)g_{\omega}(Q_t^2){\partial \Delta S^r(b=1/Q_t,Q)\over \partial Q_t^2}
\label{aa1}
\end{equation}
where the function $\Delta S^r(b=1/Q_t,Q)$ is given by equation (\ref{delsr}).  
Using equations (\ref{tg1}) and (\ref{delsr}) we get for $Q_t<Q$: 
 $$
2{\partial \bar f_{\omega}(b=1/Q_t,Q)\over \partial Q_t^2} \simeq 
$$

$$
{T_g(b=1/Q_t,Q)\over Q_t^2}\left[\int_0^{1-Q_t/Q}dz zP_{gg}(z) 
{\alpha_s\left({Q_t^2\over (1-z)^2}\right)\over 2\pi} g_{\omega}(Q_t^2)+
{d g_{\omega}(Q_t^2)\over dln(Q_t^2)}\right]+
$$
\begin{equation}
{T_g(b=1/Q_t,Q)\over Q_t^2}g_{\omega}(Q_t^2)\left\{\int_0^{1-Q_t/Q}dz 
{\alpha_s\left({Q_t^2\over (1-z)^2}\right)\over 2 \pi}
 zP_{gg}(z)[z^{\omega -1}-1]-  \int_0^{1}dz {\alpha_s(Q_t^2)\over 2 \pi}
 zP_{gg}(z)[z^{\omega -1}-1]\right\}
\label{aa2}
\end{equation}
Taking into account the DGLAP evolution equation:
\begin{equation}
\mu^2 {dg_{\omega}(\mu^2)\over d\mu^2}=
{\alpha_s(\mu^2)\over 2\pi} \int_0^1dz zP_{gg}(z)[z^{\omega-1}-1]g_{\omega}(\mu^2)
\label{aa3}
 \end{equation}
we get: 
$$
2{\partial \bar f_{\omega}(1/Q_t,Q)\over \partial Q_t^2} \simeq 
$$

\begin{equation}
{T_g(b=1/Q_t,Q)\over Q_t^2}g_{\omega}(Q_t^2)\left\{\int_0^{1-Q_t/Q}dz 
{\alpha_s\left({Q_t^2\over (1-z)^2}\right)\over 2 \pi}
 zP_{gg}(z)z^{\omega -1}\right\}
\label{aa4}
\end{equation}

Taking the inverse Mellin transform  of both sides of equation (\ref{aa4}) we get 
equation (\ref{fkmr}).

\newpage

%%%%%%%%%%%%%%%%%%%%%%%%%%%%%%%%%%%%%%%%%%%%%%%%%%%%%%%%%%%%%%%%%
\newpage 
\begin{figure}[t]   
  \vspace*{0.0cm}   
     \centerline{   
         \epsfig{figure=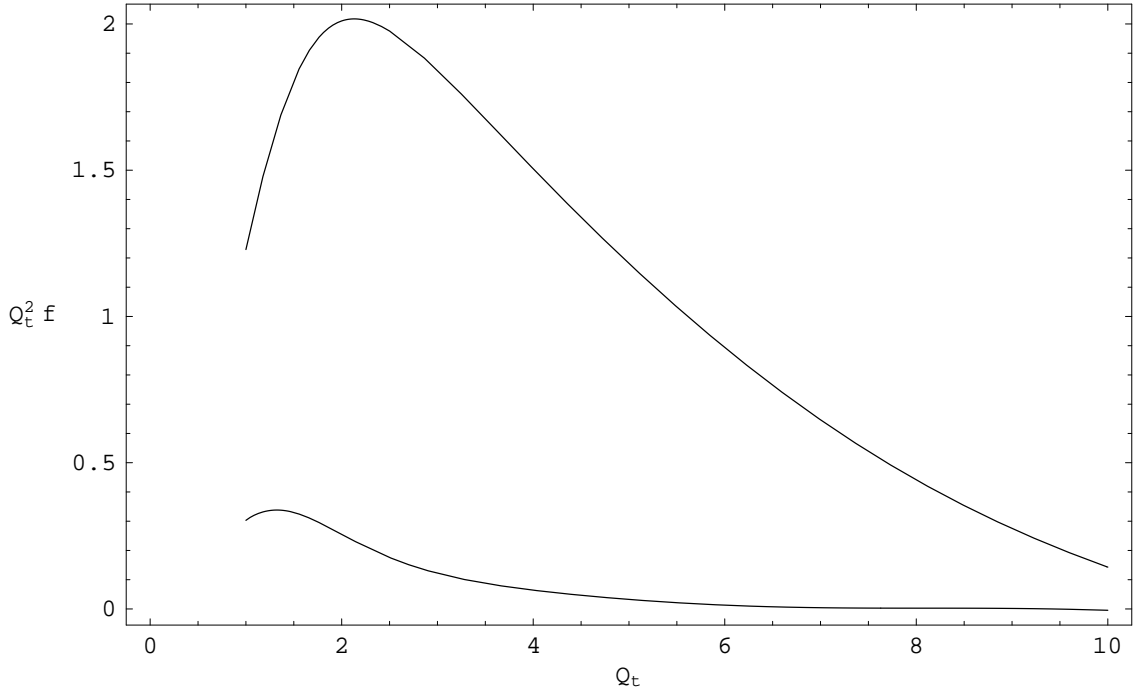,width=15cm}   
           }   
\vspace*{0.5cm}   
\caption{ 
Function $Q_t^2f(x,Q_t,Q)$, where $f(x,Q_t,Q)$ is the unintegrated gluon distribution 
obtained from the exact solution of the CCFM 
equation in the single loop approximation,  plotted as the function of 
the transverse momentum $Q_t$ of the gluon for $Q^2=100 GeV^2$. The upper and 
lower curves correspond to $x=0.01$ and $x=0.1$ 
respectively.  The transverse momentum $Q_t$ is in $GeV$. 
\label{fig:1}}   
\end{figure}  
\newpage
%%%%%%%%%%%%%%%%%%%%%%%%%%%%%%%%%%%%%%%%%%%%%%%%%%%%%%%%%%%%%%%%%%% 
\begin{figure}[t]   
  \vspace*{0.0cm}   
     \centerline{   
         \epsfig{figure=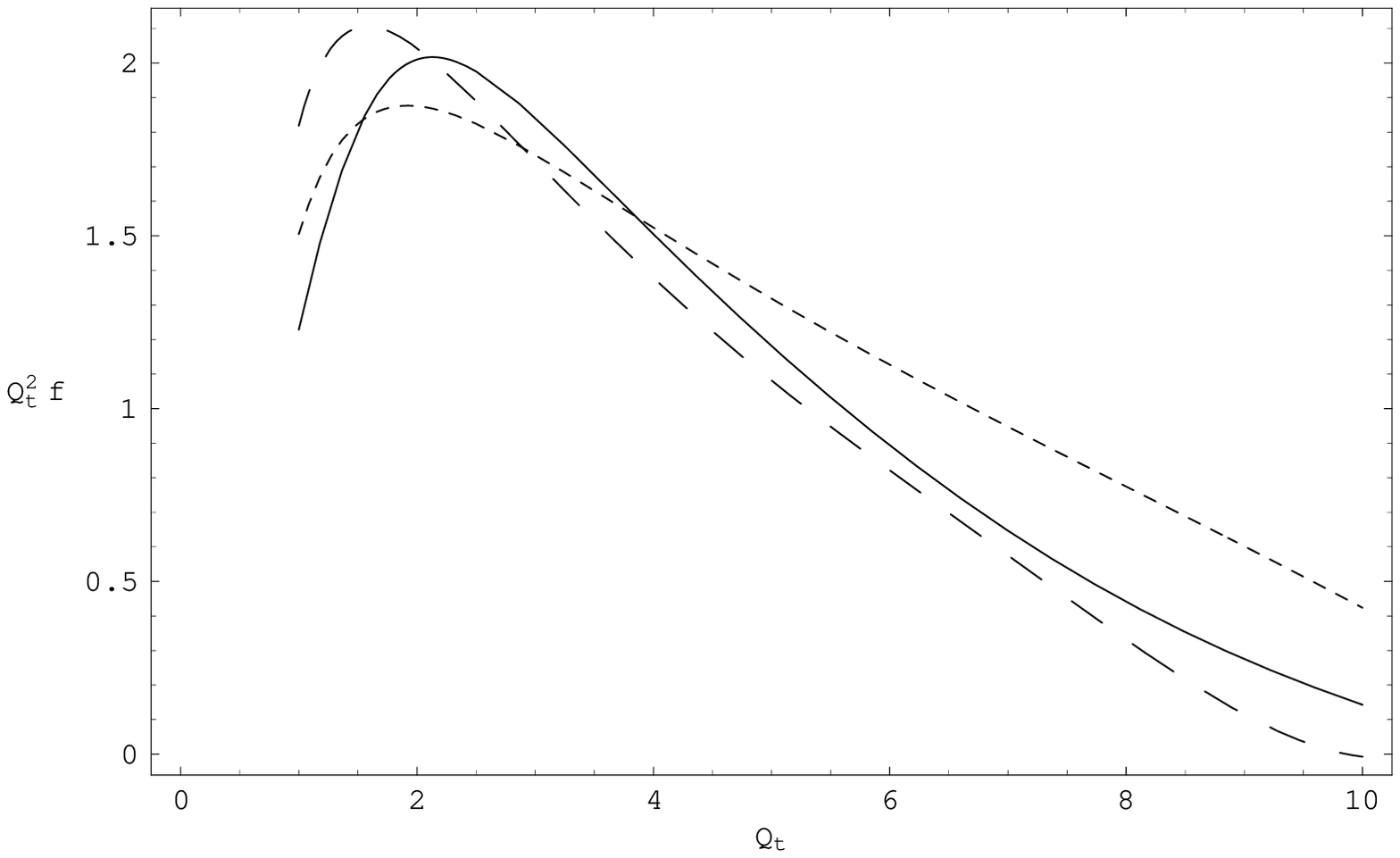,width=15cm}   
           }   
\vspace*{0.5cm}   
\caption{ 
Function $Q_t^2f(x,Q_t,Q)$, 
where $f(x,Q_t,Q)$ is the unintegrated gluon distribution 
obtained from the exact solution of the CCFM 
equation in the single loop approximation,  plotted as the function of 
the transverse momentum $Q_t$ of the gluon for $Q^2=100 GeV^2$ and $x=0.01$. 
The solid curve corresponds to $f(x,Q_t,Q)$ obtained from  exact solution of the CCFM 
equation in the single loop approximation, 
while the short dashed and long dashed curves 
correspond to approximate expressions for $f(x,Q_t,Q)$  given by equations (34) 
and (36) 
respectively.  The transverse momentum $Q_t$ is in $GeV$.   
\label{fig:2}}
\end{figure}   
%%%%%%%%%%%%%%%%%%%%%%%%%%%%%%%%%%%%%%%%%%%%%%%%%%%%%%%%%%%%%%%%%%% 
\newpage   
\begin{figure}[t]   
  \vspace*{0.0cm}   
     \centerline{   
         \epsfig{figure=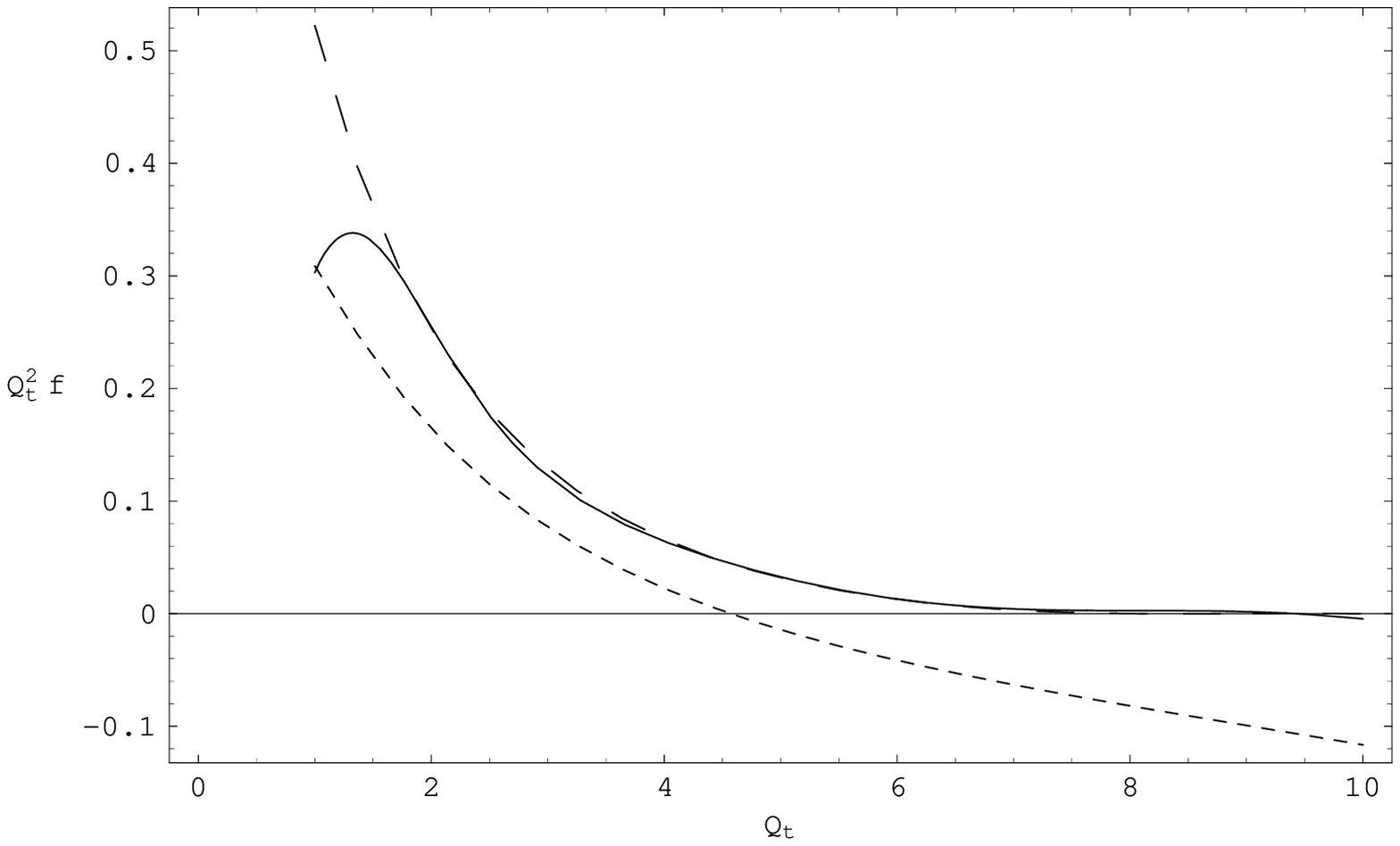,width=15cm}   
           }   
\vspace*{0.5cm}   
\caption{ Function $Q_t^2f(x,Q_t,Q)$, 
where $f(x,Q_t,Q)$ is the unintegrated gluon distribution 
obtained from the exact solution of the CCFM 
equation in the single loop approximation,  plotted as the function of 
the transverse momentum $Q_t$ of the gluon for $Q^2=100 GeV^2$ and $x=0.1$. 
The solid curve corresponds to $f(x,Q_t,Q)$ obtained from the exact solution of the CCFM 
equation in the single loop approximation, while the short dashed and long dashed curves 
correspond to approximate expressions for  $f(x,Q_t,Q)$ given by equations (34) 
and (36)  
respectively. The transverse momentum $Q_t$ is in $GeV$.    
\label{fig:3}}
\end{figure}   
%%%%%%%%%%%%%%%%%%%%%%%%%%%%%%%%%%%%%%%%%%%%%%%%%%%%%%%%%%%%%%%%%%%  


\begin{thebibliography}{9999}
\bibitem{DDT} Yu.L. Dokshitzer, D.I. Dyakonov and S.I. Troyan, Phys. Rep. {\bf 58} (1980) 269. 
\bibitem{KMR1} M.A. Kimber, A.D.Martin and M.G. Ryskin, Eur. Phys. J. {\bf C12} (2000) 
655. 
\bibitem{KMKS1}M.A. Kimber, A.D.Martin, J. Kwieci\'nski and A.M. Sta\'sto, 
Phys. Rev. {\bf D62} (2000) 094006. 
\bibitem{KMR2} M.A. Kimber, A.D.Martin and M.G. Ryskin, Phys. Rev. {\bf D63} (2001) 
114027. 
\bibitem{MR}A.D. Martin and M.G.Ryskin, Phys. Rev. {\bf D64} (2001) 094017.  
\bibitem{KHMR} V.A. Khoze, A.D. Martin and M.G. Ryskin, Eur. Phys. J. {\bf C14} (2000) 525; 
{\it ibid.} {\bf C19} (2001) 477; Erratum - {\it ibid.} {\bf C20} (2001) 599. 
\bibitem{CCFM} M. Ciafaloni, Nucl. Phys. {\bf B296} (1988) 49; 
S. Catani, F. Fiorani and G. Marchesini, Phys. Lett. {\bf B234} (1990) 339; 
Nucl. Phys. {\bf B336} (1990) 18. 
\bibitem{GM1} G. Marchesini, in Proceedings of the Workshop "QCD at 200 TeV", Erice, Italy, 
1990, edited by L. Cifarelli and Yu. L. Dokshitzer, (Plenum Press, New York, 1992), p. 183. 
\bibitem{BRW} B.R. Webber 
Nucl. Phys. B (Proc. Suppl.) {\bf 18C} (1990) 38.
\bibitem{GMBRW} G.Marchesini and B.R. Webber, Nucl. Phys. {\bf B386} (1992) 215; 
B.R. Webber in Proceedings of the Workshop "Physics at HERA", DESY, Hamburg, Germany, 1992, edited by W. Buchm\"uller and G. Ingelman 
(DESY, Hamburg, 1992). 
\bibitem{GM2}G. Marchesini, Nucl.Phys. {\bf B445} (1995) 49. 
\bibitem{KMSU}J. Kwieci\'nski, A.D. Martin and P.J. Sutton, Phys. Rev. {\bf D52} (1995) 
1445.
\bibitem{CCFMD} G. Bottazzi, G. Marchesini, G.P. Salam and M Scorletti, Nucl. Phys. {\bf 
B505} (1997) 366; JHEP {\bf 9812} (2998) 011. 
\bibitem{KGBGT}K. Golec-Biernat, L. Goerlich and J. Turnau, Nucl. Phys. {\bf B527} (1998) 
289.
\bibitem{GSAL}G.P. Salam, JHEP {\bf 9903} (1999) 009; Nucl. Phys.Proc. Suppl. {\bf 79} (1999) 
426. 
\bibitem{JUNG} H. Jung, Nucl. Phys. Proc. Suppl. {\bf 79} (1999) 429; Phys. Rev. {\bf D65} 
(2002) 034015; Comput. Phys. Commun. {\bf 143} (2002) 100. 
\bibitem{JUNGS}H. Jung and G.P. Salam, Eur. Phys. J. {\bf C19} (2001) 351. 
\bibitem{DKTM}Yu.L. Dokshitzer, V.A. Khoze, S.I. Troyan and A.H. Mueller, 
Rev. Mod. Phys. {\bf 60} (1988) 373.  
\bibitem{PT}Y.I.Dokshitzer, D.I.Dyakonov and S.I.Troyan, Phys. Lett. 
{\bf B79} (1978) 269;  G. Parisi and R.Petronzio, Nucl. Phys. {\bf B154} (1979) 427; 
J. Collins and D. Soper, Phys. Rev. {\bf D16} (1977) 2219; 
J. Collins, D. Soper, G. Sterman, Nucl. Phys. {\bf B250} (1985) 199.
\end{thebibliography}
\end{document}